\documentclass[aps,prd,amsmath,amsfonts,amssymb,eqsecnum,nofootinbib,
  floatfix,secnumarabic,preprintnumbers,superscriptaddress,nobalancelastpage,
  twocolumn]{revtex4}
\usepackage{color,graphicx}

\newcommand {\beq} {\begin{equation}}
\newcommand {\eeq} {\end{equation}}

\definecolor{red1}{cmyk}{0,1,1,0.1}
\definecolor{blue1}{cmyk}{1,0,0,0}

\newcommand{\ignore}[1]{}

\newcommand{\lhhh}{\lambda_{hhh}^{\rm SM}}
\newcommand{\lhhhh}{\lambda_{hhhh}^{\rm SM}}
\newcommand{\amc}{{\sc MadGraph5\textunderscore}a{\sc MC@NLO}}
\newcommand{\frules}{{\sc Feyn\-Rules}}

\begin{document}

\date{\today}

\title{Probing Higgs self-interactions in proton-proton collisions\\ at a center-of-mass energy of 100~TeV}

\author{Benjamin Fuks}
\email{fuks@lpthe.jussieu.fr}
\affiliation{Sorbonne Universit\'es, UPMC Univ.~Paris 06, UMR 7589, LPTHE, F-75005 Paris, France}
\affiliation{CNRS, UMR 7589, LPTHE, F-75005 Paris, France}
\author{Jeong Han Kim}
\email{jeonghan.kim@kaist.ac.kr}
\affiliation{Department of Physics, Korea Advanced Institute of Science and Technology, 335 Gwahak-ro, Yuseong-gu, Daejeon 305-701, Korea}
\affiliation{Center for Theoretical Physics of the Universe, IBS, 34051 Daejeon, Korea; and Center for Axion and Precision Physics Research, IBS, 34141 Daejeon, Korea}
\author{Seung J.~Lee}
\email{sjjlee@korea.edu}
\affiliation{Department of Physics, Korea University, Seoul 136-713, Korea}
\affiliation{School of Physics, Korea Institute for Advanced Study, Seoul 130-722, Korea}

\begin{abstract}
We present a phenomenological study of triple-Higgs production in which we
estimate the prospects for measuring the form of the Higgs potential at
future circular collider projects. We analyze proton-proton collisions at a
center-of-mass energy of 100~TeV and focus on two different signatures in which
the final state is made of four $b$-jets and either a pair of photons or a pair
of tau leptons. We study the resulting sensitivity on the Higgs cubic and
quartic self-interactions and investigate how it depends on the $b$-tagging,
tau-tagging and photon resolution performances of detectors that could be
designed for these future machines. We then discuss possible luminosity goals for
future 100~TeV collider projects that would allow for a measurement of the Higgs
potential and its possible departures from the Standard Model expectation.
\end{abstract}
\maketitle

\section{Introduction}\label{sec:intro}
The discovery in 2012~\cite{Aad:2012tfa,Chatrchyan:2012ufa} of a Higgs boson
exhibiting properties similar to those expected from the Standard Model~\cite{%
lhc:higgscouplings} has been one of the most important developments of the last
decade in experimental particle physics. Theoretically, this new state completes
the Standard Model framework and provides an explanation for both the
spontaneous breaking of the electroweak symmetry~\cite{Englert:1964et,%
Higgs:1964pj,Guralnik:1964eu} and the generation of the fermion
masses~\cite{Weinberg:1967tq}.
This observation however only consists of the first ingredient allowing one to
establish the Brout-Englert-Higgs mechanism and to fully confirm the Standard
Model nature of the observed new state. Any conclusive statement indeed
requires, in addition to the information currently available from experimental
data, at least a more detailed knowledge of the form of the Higgs potential.
Furthermore, regardless of a possible future evidence for physics beyond the
Standard Model, the Higgs potential plays a key role in our understanding of the
dynamics behind the electroweak symmetry breaking. In this context, multiple
Higgs boson probes are the simplest processes that could get sensitivity to the
Higgs trilinear and quartic self-interaction strengths, and thus to the form of
the Higgs potential. Consequently, it will receive a special attention during
the next runs of the Large Hadron Collider (LHC) and will be an important topic
of the physics program of any future high-energy machine that could be built
within the next years.

In the Standard Model, multiple Higgs production at the LHC is rather
suppressed~\cite{Glover:1987nx,Plehn:1996wb,Dawson:1998py,Frederix:2014hta,%
Maltoni:2014eza,deFlorian:2015moa,%
Grigo:2015dia}. Any precise enough direct measurement of the Higgs-boson
trilinear coupling $\lhhh$ will hence be challenging~\cite{Baur:2002rb,%
Baur:2003gpa,Baur:2003gp}, and it will be impossible to get any information on
the quartic Higgs self-interaction $\lhhhh$~\cite{Plehn:2005nk,Binoth:2006ym}.
These two coupling strengths may however differ in other theoretical frameworks,
so that the double-Higgs production channel is expected to provide valuable
information and constraints on new physics from the analysis of future collider
data~\cite{Pierce:2006dh,Dawson:1998py,Arhrib:2009hc,Asakawa:2010xj,Grober:2010yv,%
Contino:2012xk,Dolan:2012rv,Gillioz:2012se,Cao:2013si,Nhung:2013lpa,%
Ellwanger:2013ova,Liu:2013woa,No:2013wsa,Baglio:2012np,Shao:2013bz,%
Baglio:2014nea,Hespel:2014sla,Chen:2014xra,Bhattacherjee:2014bca,%
Dawson:2015oha,Grober:2015cwa}. None of
the past and present machines are nevertheless expected to get the chance of
measuring or constraining the quartic Higgs self-coupling, so that this task is
left for the experimental future of our field for which different options are
being discussed today. In this work, we explore the opportunities that are
inherent to a new accelerator facility aiming to collide highly-energetic proton
beams in the post-LHC era, and that may be built either at CERN~\cite{FCC} or at
IHEP~\cite{FCCihep}. We hence investigate triple Higgs production
in the gluon fusion channel in proton-proton
collisions at a center-of-mass energy of \mbox{$\sqrt{s}=100$~TeV} and consider
the huge statistics that could be offered by such machines aiming to collect
several tens of ab$^{-1}$ of data.

Once its decay is considered, a tri-Higgs-boson system can give rise to a
variety of final state signatures. In terms of branching ratios, the most
important channel consists of a final state made of six jets originating from
the fragmentation of $b$-quarks. Like for double Higgs-boson production when the
two Higgs bosons further decay into a four $b$-jet system, the observation of
such a signal from the overwhelming background may not be possible without the use
of either boosted object reconstruction techniques~\cite{Dolan:2012rv,%
Cooper:2013kia,deLima:2014dta} or angular information~\cite{Wardrope:2014kya}.
Since this heavily depends on the detector capabilities both in terms of jet
substructure identification and resolution, it is currently difficult to assess
the prospects of the six $b$-jet channel for pinning down a triple Higgs-boson
signal, with the detector technology to be adopted for the future proton-proton collider
projects being not decided so far. For similar reasons, it will be difficult to
determine how most of the subleading decay channels that involve $W$-boson pairs
could be used for extracting information on the Higgs quartic self-coupling~%
\cite{Papaefstathiou:2012qe,Martin-Lozano:2015dja}. We consequently consider
both a clean channel where four $b$-jets and a pair of photons ($b\bar b\ b\bar
b\ \gamma\gamma$) are issued from the Higgs decays, and a branching-ratio-%
enhanced decay mode where four $b$-jets are produced in association with a pair
of tau leptons ($b\bar b\ b\bar b\ \tau^+\tau^-$).

Our study is based on Monte Carlo simulations of proton-proton collisions at a
center-of-mass energy of 100~TeV as they could occur in the currently studied
Future Circular Collider (FCC) projects, and we generate and analyze both
background and signal events. We moreover include generic reconstruction
features and study the dependence of the FCC sensitivity to the Higgs quartic
self-coupling in terms of different goals for the detector performances. In
particular, we investigate the robustness of our findings in terms of
$b$-tagging and $\tau$-tagging efficiencies and mistagging rates, as well as
in terms of photon reconstruction properties. We extend in this way previous
studies that appeared at the time of the write-up of this paper~\cite{%
Papaefstathiou:2015paa,Chen:2015gva}, and we study the consequences of variations from
an optimal search strategy on the FCC sensitivity to the quartic Higgs
coupling.

The rest of this paper is organized as follows. In Section~\ref{sec:model}, we
discuss the theoretical framework that we have adopted in our study and present
details on the Monte Carlo simulations that we have performed.
Section~\ref{sec:results} explores the prospects for the measurement of the
Higgs potential by analyzing two signatures of a triple-Higgs signal, namely
channels with a final state featuring four $b$-jets and either a pair of photons
or a pair of tau leptons. We then study the FCC sensitivity
to deviations from the Standard Model in the Higgs trilinear and
quartic interactions, and finally discuss our conclusions in Section~\ref{%
sec:conclusions} where we also investigate the FCC luminosity goals
that should be aimed for in order to access the Higgs self-interactions.

\section{Theoretical framework and Monte Carlo simulation details}\label{sec:model}
\subsection{Triple Higgs boson production and decay}
\label{sec:theo}
Our phenomenological analysis of the sensitivity of the FCC to triple
Higgs-boson events relies on Monte Carlo simulations of proton-proton collisions
to be produced at a center-of-mass energy of \mbox{$\sqrt{s}=100$~TeV}. To this
aim, we generate partonic events associated with the loop-induced
\mbox{$g g\to hhh$} subprocess within the \amc\ framework~\cite{Alwall:2014hca}
that has been recently extended to handle
loop-induced processes~\cite{Hirschi:2015iia}. Our
theoretical model description is based on the Standard Model after having
modified the Higgs potential to allow for deviations induced by new physics. We
parameterize the latter in a model-independent fashion,
\beq
  V_{\rm h} =  \frac{m_h^2}{2} h^2 + (1 + \kappa_3) \lhhh v h^3 +
    \frac14 (1 + \kappa_4) \lhhhh h^4 \ ,
\label{eq:vh}\eeq
where in our notation, $h$ denotes the physical Higgs boson field, $v$ stands
for its related vacuum expectation value, $m_h$ for its mass, and the Standard
Model self-interaction strengths are given by
\beq
  \lhhh = \lhhhh = \frac{m_h^2}{2 v^2} \ .
\eeq
Following the strategy of Ref.~\cite{Christensen:2009jx}, we have implemented
the above modifications of the scalar potential within the Standard Model
implementation shipped with the
\frules~package~\cite{Alloul:2013bka} and made use of the {\sc NloCT}
program~\cite{Degrande:2014vpa} to generate a UFO library~\cite{Degrande:2011ua}
containing both tree-level and loop-level information. In particular, this UFO
module includes the $R_2$ counterterms relevant for the evaluation of the loop
integrals in four dimensions, as performed in \amc~\cite{Hirschi:2011pa} which
follows the Ossola-Papadopoulos-Pittau formalism~\cite{Ossola:2006us,%
Ossola:2008xq}.

\begin{figure}
\includegraphics[width=\columnwidth]{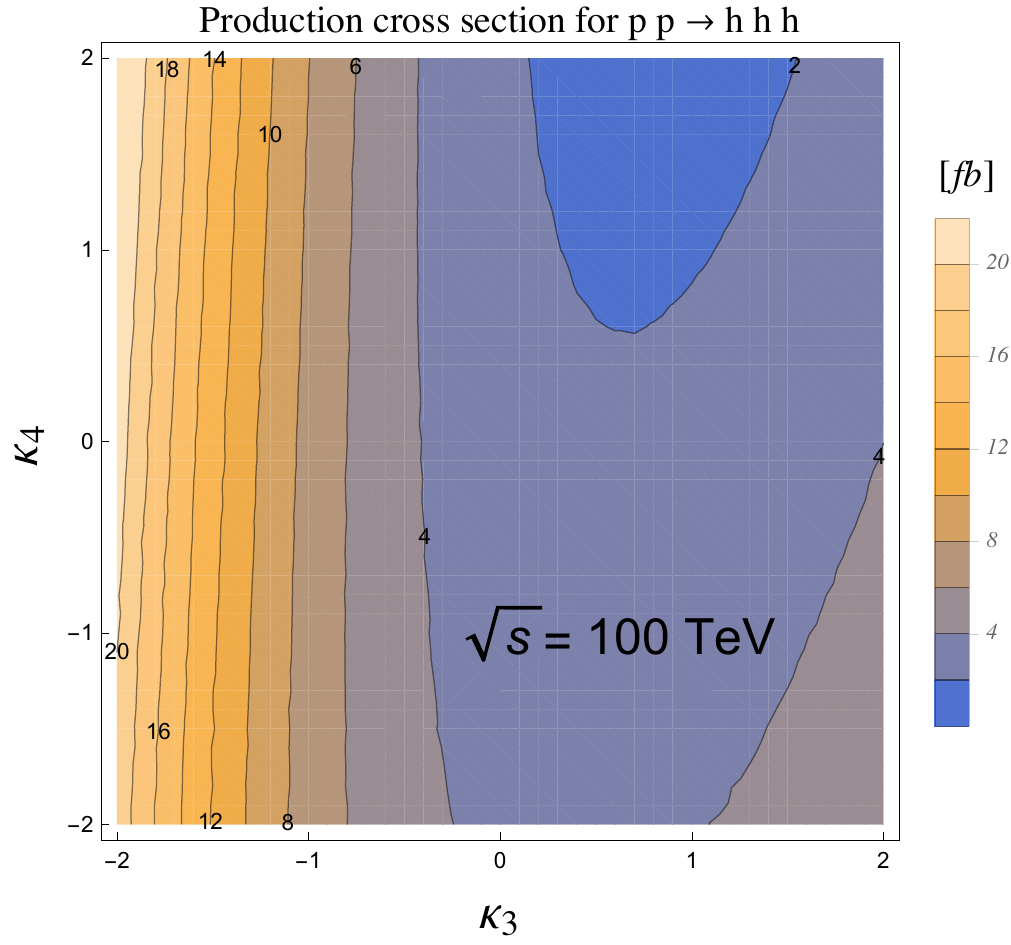}
\caption{Leading-order total cross section for triple Higgs-boson production
 from proton-proton collisions at \mbox{$\sqrt{s}=100$~TeV}. The results are
 presented in terms of the $\kappa_3$ and $\kappa_4$ parameters defined in
 Eq.~\eqref{eq:vh}.}
\label{fig:k3_k4_Xsection}
\end{figure}

We calculate the triple Higgs-boson production cross section $\sigma_{\rm hhh}$
at the leading-order accuracy by convoluting the one-loop
squared matrix elements generated by \amc\ with the leading-order set of
NNPDF~2.3 parton distribution functions~\cite{Ball:2012cx}. We present the
results in Figure~\ref{fig:k3_k4_Xsection} where we show the variation of
$\sigma_{\rm hhh}$ in the $(\kappa_3, \kappa_4)$ plane. We first observe that
$\sigma_{\rm hhh}$ is very sensitive to independent modifications of the
trilinear Higgs coupling $\kappa_3$, and more specifically when $\kappa_3$ has
a negative value. The dependence on the $\kappa_4$ parameter is milder,
although large variations can be seen once the value of $\kappa_3$ is fixed. We
will establish, in the next section, how the FCC could be sensitive to such
variations and constrain the $(\kappa_3, \kappa_4)$ parameter space.

Triple Higgs production leads to a large class of possible final state
signatures once the Higgs-boson decays into a $b\bar b$ pair (with a branching
ratio of 0.58), a $WW^\ast$ pair (with a branching ratio of 0.22), a
$\tau^+\tau^-$ pair (with a branching ratio of 0.064), a $ZZ^\ast$ pair (with a
branching ratio of 0.027) and into a $\gamma\gamma$ pair (with a branching ratio of
0.0023) are accounted for. The dominant channel corresponds to a final state
comprised of six $b$-jets, with an associated branching ratio of 19.5\%. Once a
semi-realistic $b$-tagging efficiency of
70\% is included, this number drops to 14\% (2.3\%) when we require at least
four (exactly six) $b$-tagged jets. The observation of such a tri-Higgs signal
may however be complicated in particular due to the multijet background, and
advanced analysis techniques may have to be used, as for the case of di-Higgs
production at the LHC~\cite{Dolan:2012rv,Cooper:2013kia,deLima:2014dta,%
Wardrope:2014kya}. Such techniques are however strongly tied to the details of
the detectors, such as tracking performance and calorimetric granularity. We
therefore leave the study of this channel as an open question and focus instead
on the analysis of final state topologies that could be performed with any
conceivable detector design.

The next-to-dominant channel concerns the decay of the triple-Higgs system into
four $b$-jets and a pair of $W$-bosons, at least one of them being off-shell.
After imposing the semileptonic decay of the $W$-boson pair and including
$b$-tagging efficiencies, the corresponding effective branching ratio reaches
1.5\%. For the same reasons as those mentioned in the six $b$-jet case,
cornering such a triple-Higgs boson signal within the background may require
the use of techniques relying on the exact knowledge of the detector
performances, as it is already the case for di-Higgs
production at the LHC~\cite{Papaefstathiou:2012qe,Martin-Lozano:2015dja}.
We therefore ignore such a channel in our analysis, together with
any other decay mode involving weak bosons.

As a consequence, we focus on the $b\bar{b}b\bar{b}\gamma \gamma$
and $b\bar{b}b\bar{b}\tau^+ \tau^-$ decay channels. They feature small
branching ratios of 0.232\% and 6.46\% respectively, but offer
good hopes to be observable even after accounting for $b$-tagging and
tau-tagging efficiencies, in particular as the FCC luminosity goal is of
several tens of ab$^{-1}$~\cite{FCC,FCCihep}. A significant number of
decayed triple-Higgs events is thus expected to be produced.
As all other decay modes of the triple-Higgs boson system imply much smaller branching
ratios, they will be ignored in our analysis.

\subsection{Event generation and analysis methods}
\label{sec:EventGeneration}

For the simulation of the signal and background processes, we make use of the
model implementation described in Section~\ref{sec:theo} and the \amc\
framework, and generate events at the leading-order accuracy in QCD. We
additionally normalize all background samples by multiplying the leading-order
rates by a conservative $K$-factor of 2. QCD corrections to the signal process,
that are known to be large~\cite{Maltoni:2014eza}, are not included so that the
results presented below can be seen as conservative. At the generation level,
we require all
produced final-state particles to have a transverse-momentum $p_T > 15$~GeV, a
pseudorapidity satisfying $|\eta|< 5$ and to be separated from each other by
an angular distance, in the transverse plane, of $\Delta R > 0.4$.

\begin{figure}
\includegraphics[scale=0.8]{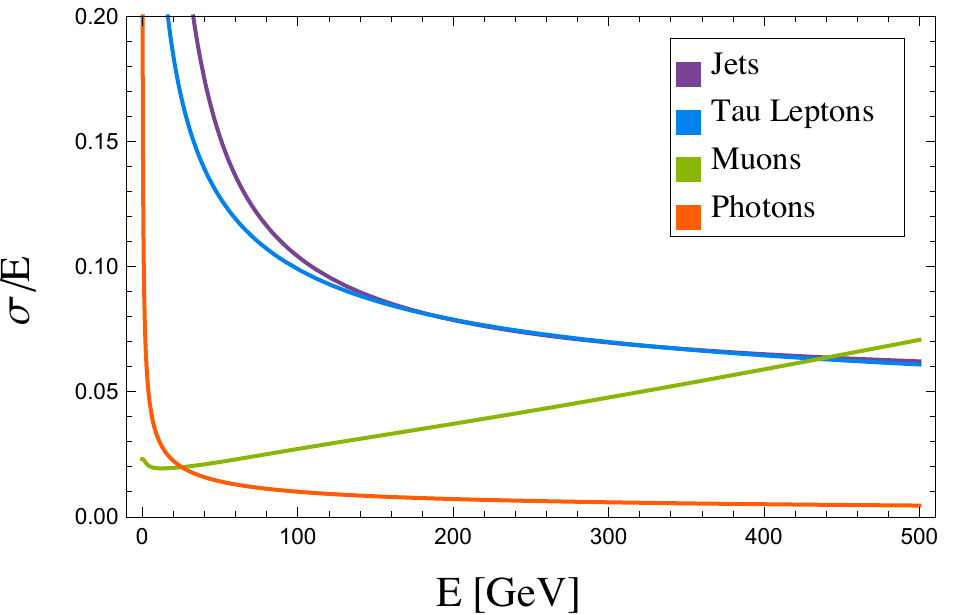}
\caption{Relative effective resolution of the different objects used in our
analysis presented as a function of their transverse-momentum.}
\label{fig:resolution}
\end{figure}

We perform our analysis at the partonic-level, therefore neglecting the possible
impact of the parton showering and the hadronization. We however include
simplistic detector effects based on the ATLAS detector performances and smear
the momentum and energy of the produced photons~\cite{Aad:2014nim}, jets and
taus~\cite{ATLAS:2013004} according to the value of their transverse momentum.
We summarize the $p_T$-dependence of the resolution functions that we have
employed in Figure~\ref{fig:resolution}. The figure shows in particular that
photons can be remarkably well reconstructed, with a relative effective
resolution of $\sigma / E \sim 0.1 / \sqrt{E}$ that only weakly depends on the
energy. We consequently expect that the reconstruction of the Higgs mass from a
diphoton system will result in a relatively narrow peak visible in the diphoton
invariant-mass spectrum and centered on the true Higgs-boson mass value of
125~GeV.

Our analysis heavily relies on $b$-jet identification as both considered search
channels contain four (parton-level) $b$-jets, while the investigation of the
$b\bar b b \bar b \tau^+\tau^-$ channel additionally depends on the efficiency
of the $\tau$-tagger. One of the aim of our study is to assess the sensitivity
reach of the FCC in the $(\kappa_3,\kappa_4)$ plane for several $b$-tagging and
$\tau$-tagging performances so that our results could be used as benchmarks for
the FCC detector design. We consider two $b$-tagging setups, with an
efficiency of 70\%/60\% for a mistagging rate of a $c$-jet as a $b$-jet of
18\%/1.8\% and of a lighter jet as a $b$-jet of 1\%/0.1\%~\cite{ATLAS:2014014}.
We then investigate the outcome of an optimistic $\tau$-tagging efficiency of
80\% whose associated mistagging rate of a jet as a tau is given by
0.1\%~\cite{Dolan:2012rv}, and also make use of a conservative $\tau$-tagging
efficiency of 50\% for a fake rate of 1\%.

\section{Phenomenological investigations}\label{sec:results}
We perform a study of a triple-Higgs signal produced in proton-proton collisions
at a center-of-mass energy of 100~TeV and show how it can be observed above the
Standard Model background in two different channels. We analyze final states
comprised either of four $b$-jets and a pair of photons (Section~\ref{Sec2a4b}),
or of four $b$-jets and a pair of tau leptons (Section~\ref{Sec2tau4b}).

\subsection{The $h h h \to \gamma \gamma b \bar{b} b \bar{b}$ final state}
\label{Sec2a4b}

We focus first on a subprocess in which the triple-Higgs-boson system decays
into four $b$-jets and a photon pair. Our analysis strategy relies mainly on the
two photons which consist of a clean probe for new physics as it is associated
with a small Standard Model background. Although the diphoton component of the
final state could provide an efficient handle for background rejection and
signal detection, the considered process suffers from a significant
reduction of the production cross section due to the small branching fraction of
a Higgs boson into a photon pair.

\begin{figure}

\includegraphics[scale=0.8]{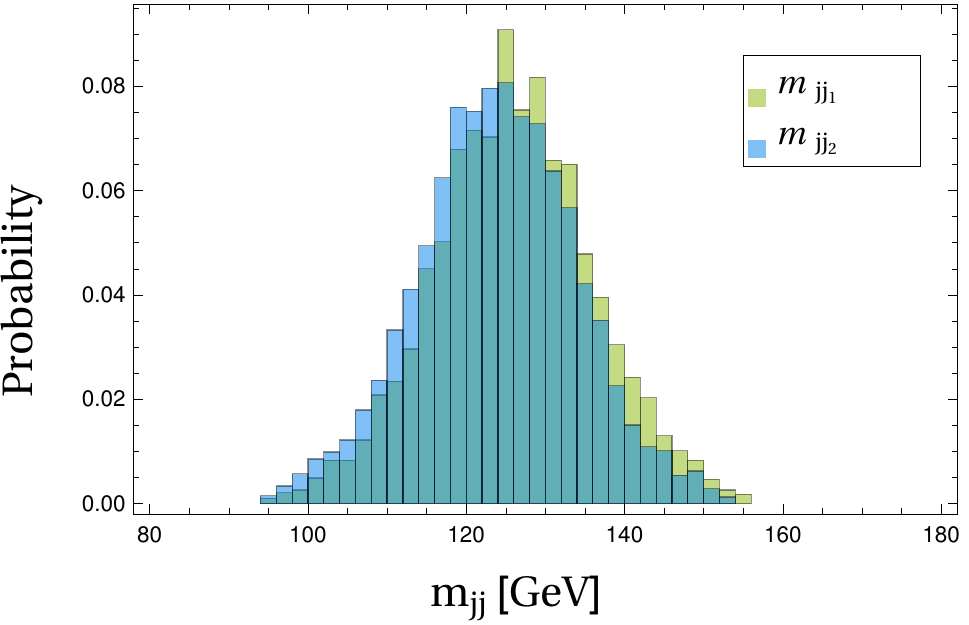} \\[.2cm]
\includegraphics[scale=0.8]{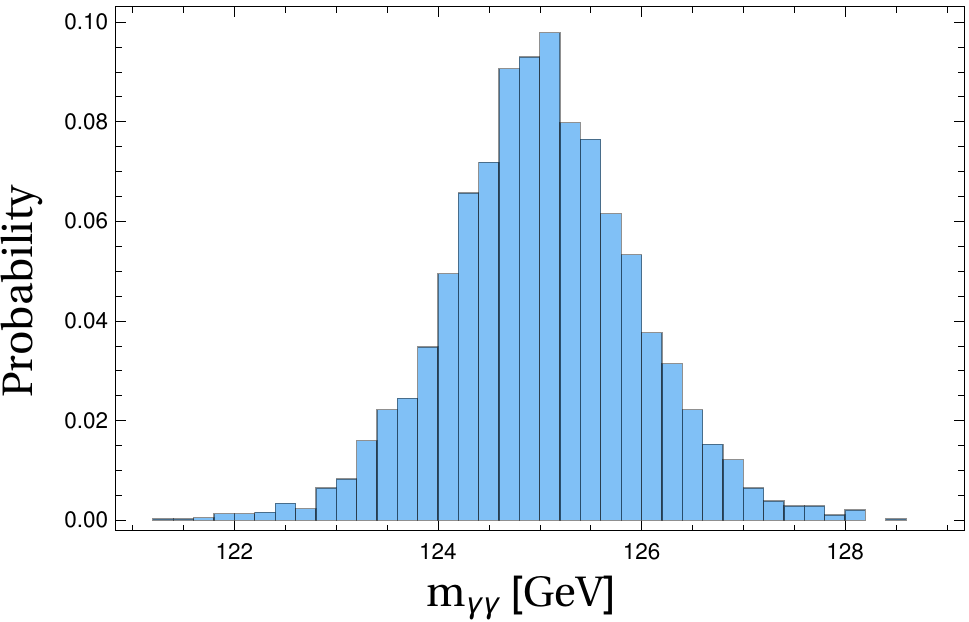}
\caption{Invariant-mass distributions of the three reconstructed Higgs bosons
from the four final-state jets (upper panel) and two final-state photons (lower
panel) after applying the preselection. We have considered a benchmark scenario
in which $\kappa_3 = -2$ and $\kappa_4 = 0$.}
\label{fig:Mh_4b_2a}
\end{figure}

On the basis of the final state topology, events are preselected by demanding
that they contain at least four jets with a transverse momentum $p_T$ and a
pseudorapidity $\eta$ satisfying \mbox{$p_T^{\rm j_1} > 50$~GeV},
\mbox{$p_T^{\rm j_2} > 30$~GeV}, \mbox{$p_T^{\rm j_3} > 20$~GeV},
\mbox{$p_T^{\rm j_4} > 15$~GeV} and \mbox{$|\eta^{\rm j_i} |< 2.5$} for
\mbox{$i=1,2,3,4$}. In addition, we require the presence of two photon
candidates whose transverse momentum and pseudorapidity fulfill
\mbox{$p_T^{\rm \gamma_1} > 35$~GeV}, \mbox{$p_T^{\rm \gamma_2} > 15$~GeV} and
$|\eta^{\rm \gamma_j} |< 2.5$ for \mbox{$j=1,2$}. In order to reduce a
possible signal contamination by jets misidentified as photons, we impose that
the photons are isolated in a way in which the transverse energy $E_{\rm T,iso}$
lying in a cone of radius \mbox{$R_{\rm iso}=0.3$} centered on each photon is
smaller than 6~GeV~\cite{ATLAS:2014iha}.

We then reconstruct the two Higgs bosons originating from the four jets
and impose that their invariant masses $m_{jj_1}$ and $m_{jj_2}$ satisfy
\mbox{$|m_h - m_{\rm j j_k}| < 15$~GeV} for \mbox{$k=1,2$}. In cases where
there are more than one combination of dijet systems compatible with this
criterion, we select the one minimizing the mass asymmetry
\begin{equation}
  \Delta_{\rm j j_1, j j_2} = \frac{m_{\rm j j_1} - m_{\rm j j_2}}{ m_{\rm j j_1} + m_{\rm j j_2} } \; .
\label{MassAsymmetry}
\end{equation}
The remaining Higgs boson is reconstructed from the diphoton system and
we demand that its invariant mass $m_{\gamma\gamma}$ fulfills
\mbox{$|m_h-m_{\gamma\gamma}| < M$} where the threshold $M$ can vary from 1
to 5~GeV. Illustrative signal distributions for a benchmark scenario in which
$\kappa_3=-2$ and $\kappa_4=0$ are shown on Figure~\ref{fig:Mh_4b_2a}. The
results however only mildly depend on the choices for the $\kappa$ parameters. We
therefore already conclude at that stage of our analysis that a high-quality
mass resolution in the diphoton spectrum will be an incontrovertible ingredient
to be able to reasonably disentangle a signal from the background. We finally require
that the selected events feature at least $N_b^{\rm min}$ $b$-tagged jets with
$N_b^{\rm min}=2$, 3 or 4.

\begin{figure}
\includegraphics[scale=0.8]{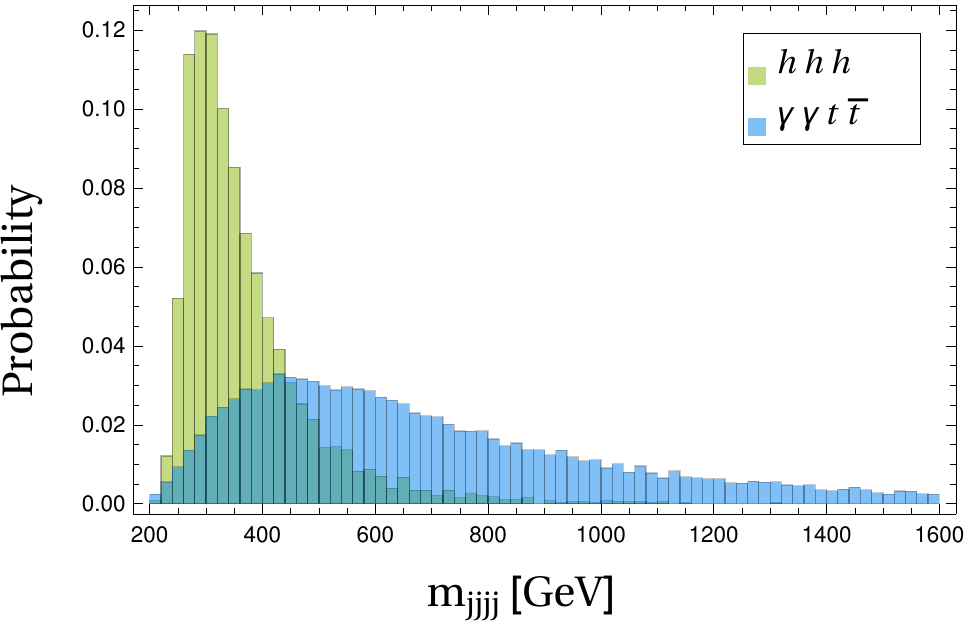}
\caption{Invariant-mass distributions of the four leading jets for both the
triple-Higgs signal and the $t\bar t\gamma\gamma$ background. Event
preselection has been applied and we have considered a scenario in which
$\kappa_3 = -2$ and $\kappa_4 = 0$.}
\label{fig:Mass4b}
\end{figure}

\begin{figure}
\includegraphics[scale=0.8]{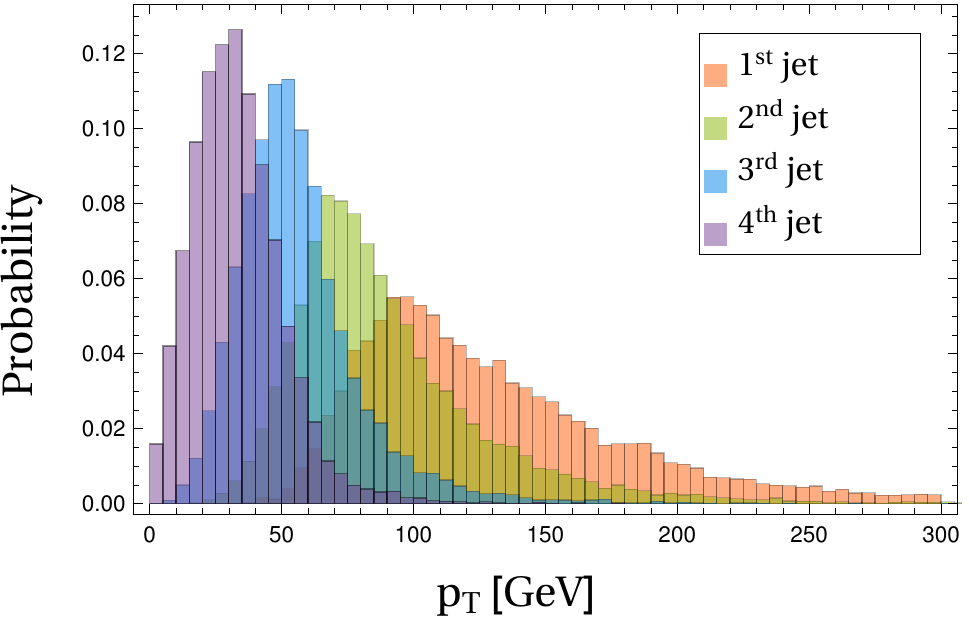} \\[.2cm]
\includegraphics[scale=0.8]{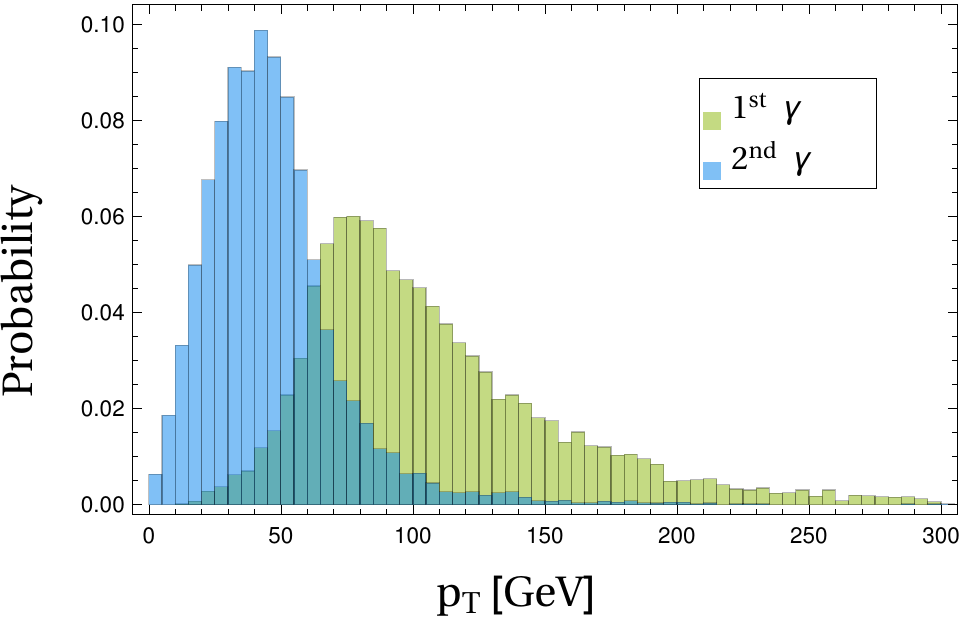}
\caption{Transverse-momentum distributions of the four $b$-jets and of the two photons
arising from a triple-Higgs signal before applying any event selection and for a
scenario in which \mbox{$\kappa_3 = -2$} and $\kappa_4 = 0$.}
\label{fig:pT_4b_2a}
\end{figure}

After this selection, the dominant sources of Standard Model background
consist of $\gamma\gamma b\bar{b}jj$, $\gamma\gamma t\bar{t}$ (with both top
quarks decaying hadronically), $\gamma\gamma Z_{bb}jj$,
$h_{\gamma\gamma}h_{bb}Z_{bb}$ and $h_{\gamma\gamma} b\bar{b}b\bar{b}$ events,
with $h_{XX}$ and $Z_{XX}$ indicating a Higgs and a $Z$-boson decaying into an
$XX$ final state respectively. All other Standard Model processes, including in
particular $\gamma\gamma b\bar{b}b\bar{b}$ and $\gamma \gamma Z_{bb} Z_{bb}$
production, have been found to yield a negligible impact.
It is further possible to reduce the $\gamma \gamma t\bar{t}$ background by
constraining the invariant-mass of the four-jet system $m_{\rm jjjj}$. In the
background case, typical $m_{\rm jjjj}$ values tend to be large since the decay
products of a massive top quark can acquire a significant $p_T$. This contrasts
with the signal, as depicted on Figure~\ref{fig:Mass4b} for a setup in which
\mbox{$\kappa_3 = -2$} and \mbox{$\kappa_4 = 0$}. Constraining $m_{\rm jjjj}$ to
be small could hence reduce the background and maintain a good signal efficiency. This
property is further illustrated in Figure~\ref{fig:pT_4b_2a} where we show the
transverse-momentum spectra of the four leading jets in the signal case. The
resulting invariant-mass distribution consequently peaks at a low value since
the jets have most of the time a small $p_T$. We have verified that
this feature is independent of the values of the $\kappa$ parameters, and
have found that requiring \mbox{$m_{\rm jjjj} < 600$~GeV} significantly reduces
the $\gamma\gamma t\bar{t}$ background without affecting the signal.

\begin{table*}
\setlength{\tabcolsep}{2.5mm}
\renewcommand{\arraystretch}{1.2}
\begin{tabular}{c|cccccc|c}
  Selection step & Signal & $\gamma\gamma b\bar{b}jj$ & $\gamma\gamma Z_{bb}jj$
     & $\gamma\gamma t\bar{t}$ & $h_{\gamma \gamma} h_{bb} Z_{bb}$
     & $h_{\gamma\gamma} b\bar{b}b\bar{b}$  &  $\sigma $ \\
  \hline \hline
  Preselection  & 19~ab & $4.2 \times 10^{6}$~ab & $5.3 \times 10^{4}$~ab
      & $1.1 \times 10^{5}$~ab & 0.990~ab & 7.10~ab & 0.04\\
  $| m_h - m_{\rm j j_1, j j_2}| < 15$~GeV & 14~ab & $1.7 \times 10^{5}$~ab
      & $1.8 \times 10^{3}$~ab & $1.1 \times 10^{4}$~ab & 0.059~ab & 0.29~ab
      & 0.15\\
  \hline
  $| m_h - m_{ \gamma \gamma}| < 5$~GeV & 14~ab & $6.9 \times 10^{3}$~ab & 68~ab
      & 500~ab & 0.058~ab & 0.29~ab & 0.75\\
  $m_{\rm jjjj} < 600$~GeV & 14~ab & $6.9 \times 10^{3}$~ab & 68~ab & 280~ab
      & 0.058~ab & 0.29~ab & 0.72\\
  At least 2 $b$-tagged jets & 8.9~ab & 850~ab & 19~ab & 47~ab & 0.037~ab
      & 0.19~ab & 1.3\\
  At least 3 $b$-tagged jets & 6.5~ab & 12~ab & 0.26~ab & 0.89~ab & 0.027~ab
      & 0.14~ab &  6.9\\
  At least 4 $b$-tagged jets & 1.8~ab & $9.6 \times 10^{-3}$~ab
      & $2.0 \times 10^{-3}$~ab & $1.9 \times 10^{-3}$~ab
      & $7.5 \times 10^{-3}$~ab & 0.038~ab & 7.9\\
  \hline
  $| m_h - m_{ \gamma \gamma}| < 2$~GeV & 14~ab & $2.9 \times 10^{3}$~ab & 34~ab
      & 210~ab & 0.056~ab & 0.28~ab & 1.1\\
  $m_{\rm jjjj} < 600$~GeV & 13~ab & $2.9 \times 10^{3}$~ab &34~ab & 120~ab
      & 0.055~ab & 0.28~ab & 1.1\\
  At least 2 $b$-tagged jets & 8.6~ab & 630~ab &12~ab & 15~ab& 0.036~ab
      & 0.18~ab &1.5\\
  At least 3 $b$-tagged jets & 6.3~ab& 5.6~ab &0.025~ab & 0.38~ab & 0.026~ab
      & 0.13~ab &8.8\\
  At least 4 $b$-tagged jets & 1.7~ab &$4.6 \times 10^{-3}$~ab
      & $1.2 \times 10^{-5}$~ab  & $1.1 \times 10^{-3}$~ab
      & $7.1 \times 10^{-3}$~ab  & 0.036~ab & 7.8\\
  \hline
  $| m_h - m_{ \gamma \gamma}| < 1$~GeV & 11~ab &$1.2 \times 10^{3}$~ab &34~ab
      & 94~ab & 0.041~ab & 0.22~ab& 1.3\\
  $m_{\rm jjjj} < 600$~GeV& 10~ab &$1.2 \times 10^{3}$~ab &34~ab & 54~ab
     & 0.040~ab & 0.22~ab& 1.3\\
  At least 2 $b$-tagged jets & 6.5~ab & 420~ab&12~ab& 12~ab & 0.026~ab& 0.14~ab
     &1.4\\
  At least 3 $b$-tagged jets& 4.8~ab   & 4.4~ab &0.025~ab & 0.12~ab & 0.019~ab
     & 0.10~ab & 7.7\\
  At least 4 $b$-tagged jets & 1.3~ab&$3.9 \times 10^{-3}$~ab
     &$1.2 \times 10^{-5}$~ab & $4.8 \times 10^{-4}$~ab & $5.2 \times10^{-3}$~ab
      & 0.029~ab &6.8\\
\end{tabular}\\[.2cm]

\setlength{\tabcolsep}{2.5mm}
\renewcommand{\arraystretch}{1.2}
\begin{tabular}{c|cccccc|c}
  Selection step & Signal & $\gamma\gamma b\bar{b}jj$ & $\gamma\gamma Z_{bb}jj$
     & $\gamma\gamma t\bar{t}$ & $h_{\gamma \gamma} h_{bb} Z_{bb}$
     & $h_{\gamma\gamma} b\bar{b}b\bar{b}$  &  $\sigma $ \\
  \hline \hline
  Preselection  & 19~ab & $4.2 \times 10^{6}$~ab & $5.3 \times 10^{4}$~ab
      & $1.1 \times 10^{5}$~ab & 0.990~ab & 7.10~ab & 0.04\\
  $| m_h - m_{\rm j j_1, j j_2}| < 15$~GeV & 14~ab & $1.7 \times 10^{5}$~ab
      & $1.8 \times 10^{3}$~ab & $1.1 \times 10^{4}$~ab & 0.059~ab & 0.29~ab
      & 0.15\\
  \hline
  $| m_h - m_{ \gamma \gamma}| < 5$~GeV & 14~ab & $6.9 \times 10^{3}$~ab & 68~ab
      & 500~ab & 0.058~ab & 0.29~ab & 0.75\\
  $m_{\rm jjjj} < 600$~GeV & 14~ab & $6.9 \times 10^{3}$~ab & 68~ab & 280~ab
      & 0.058~ab & 0.29~ab & 0.72\\
  at least 2 $b$-tagged jets & 11~ab & $1.3 \times 10^{3}$~ab & 27~ab & 74~ab
      & 0.045~ab & 0.23~ab & 1.3\\
  at least 3 $b$-tagged jets & 8.9~ab  & 160~ab & 3.5~ab& 12~ab & 0.038~ab
      & 0.19~ab  &2.9\\
  at least 4 $b$-tagged jets & 3.3~ab & 1.3~ab & 0.27~ab & 0.26~ab & 0.014~ab
      & 0.071~ab & 7.4\\
  \hline
  $| m_h - m_{ \gamma \gamma}| < 2$~GeV & 14~ab & $2.9 \times 10^{3}$~ab & 34~ab
      & 210~ab & 0.056~ab & 0.28~ab & 1.1\\
  $m_{\rm jjjj} < 600$~GeV & 13~ab & $2.9 \times 10^{3}$~ab &34~ab & 120~ab
      & 0.055~ab & 0.28~ab & 1.1\\
  at least 2 $b$-tagged jets & 10~ab & 890~ab &17~ab& 25~ab & 0.043~ab &0.22~ab
      &  1.5\\
  at least 3 $b$-tagged jets & 8.6~ab & 76~ab &0.33~ab & 5.2~ab & 0.036~ab
      & 0.18~ab  & 4.1 \\
  at least 4 $b$-tagged jets & 3.2~ab &0.62~ab &$1.7 \times 10^{-3}$~ab 
      & 0.15~ab & 0.013~ab &0.067~ab  &8.6 \\
   \hline
  $| m_h - m_{ \gamma \gamma}| < 1$~GeV & 11~ab &$1.2 \times 10^{3}$~ab &34~ab
      & 94~ab & 0.041~ab & 0.22~ab& 1.3\\
  $m_{\rm jjjj} < 600$~GeV& 10~ab &$1.2 \times 10^{3}$~ab &34~ab & 54~ab
     & 0.040~ab & 0.22~ab& 1.3\\
   at least 2 $b$-tagged jets & 7.9~ab & 590~ab &17~ab & 17~ab & 0.031~ab
     & 0.17~ab  &1.4\\
   at least 3 $b$-tagged jets & 6.6~ab & 59~ab &0.33~ab & 1.7~ab & 0.026~ab
     & 0.14~ab&3.6\\
   at least 4 $b$-tagged jets & 2.4~ab &0.54~ab &$1.7 \times 10^{-3}$~ab
     & 0.065~ab& $9.6 \times 10^{-3}$~ab  & 0.053~ab  & 7.5\\
\hline
\end{tabular}
\caption{Effects of our selection strategy for an illustrative benchmark
  scenario in which \mbox{$\kappa_3=-2$} and \mbox{$\kappa_4=0$}. We show the
  resulting cross sections after each of the selection steps. In the upper
  (lower) table, we assume a $b$-tagging efficiency of $60\%$ ($70\%$) and a
  mistagging rate of $c$ and lighter jets as a $b$-jet of $1.8\%$ ($18\%$) and
  $0.1\%$ ($1\%$) respectively. The significance $\sigma$ is
  calculated for a luminosity of 20~ab$^{-1}$.}
\label{tab:aCutflow}
\end{table*}

The $b$-tagging strategy plays a central role in the possibility of observing a
signal from the background. We start by making use of a conservative estimate
for the $b$-tagging performances, and consider a tagging efficiency of $60\%$ for
a mistagging rate of $1.8\%$ and $0.1\%$ for $c$ and lighter jets respectively.
We present in Table~\ref{tab:aCutflow} (upper table) the effects of
our selection strategy for a benchmark scenario in which \mbox{$\kappa_3=-2$}
and \mbox{$\kappa_4=0$} and for a given luminosity of 20~ab$^{-1}$. We have
varied the diphoton invariant-mass resolution $M$ from 1~GeV to 5~GeV and the
minimum number of required $b$-tagged jets from $N_b^{\rm min}=2$ to
$N_b^{\rm min}=4$. We have found that for all choices of $M$ values, a demand of
at least either three or four $b$-tagged jets is in order so that one could get
some sensitivity to the signal. We have here defined the significance $\sigma$ as a
likelihood ratio~\cite{Cowan:2010js},
\beq
  \sigma \equiv \sqrt{-2\, \ln\frac{L(S + B|B)}{L(B|B)} }\, ,
\label{eq:sign}\eeq
where $S$ and $B$ represent the number of selected signal and
background events respectively, and $L$ is the likelihood function
\beq
  L(x |n) =  \frac{x^{n}}{n !} e^{-x} \,.
\eeq
For this very specific scenario, it is not possible to determine the diphoton
mass resolution that would be necessary for observing the signal.

\begin{figure*}
\includegraphics[scale=0.8]{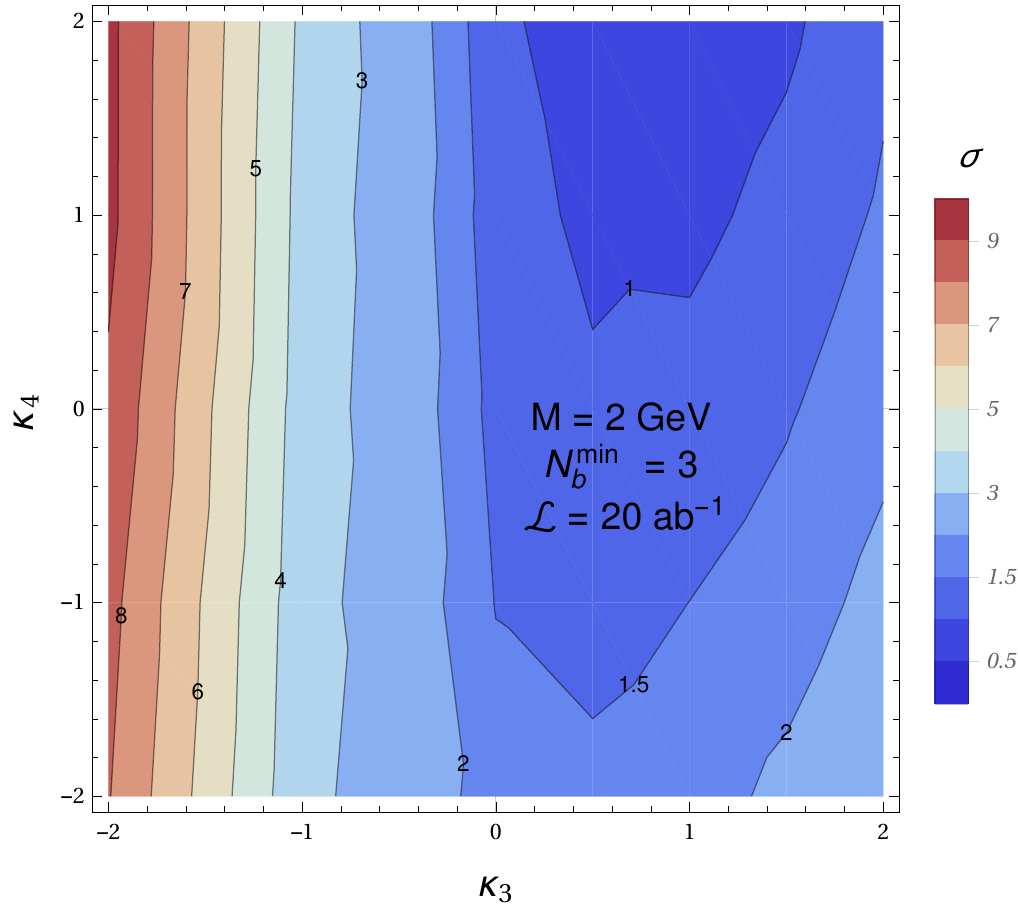}
\includegraphics[scale=0.8]{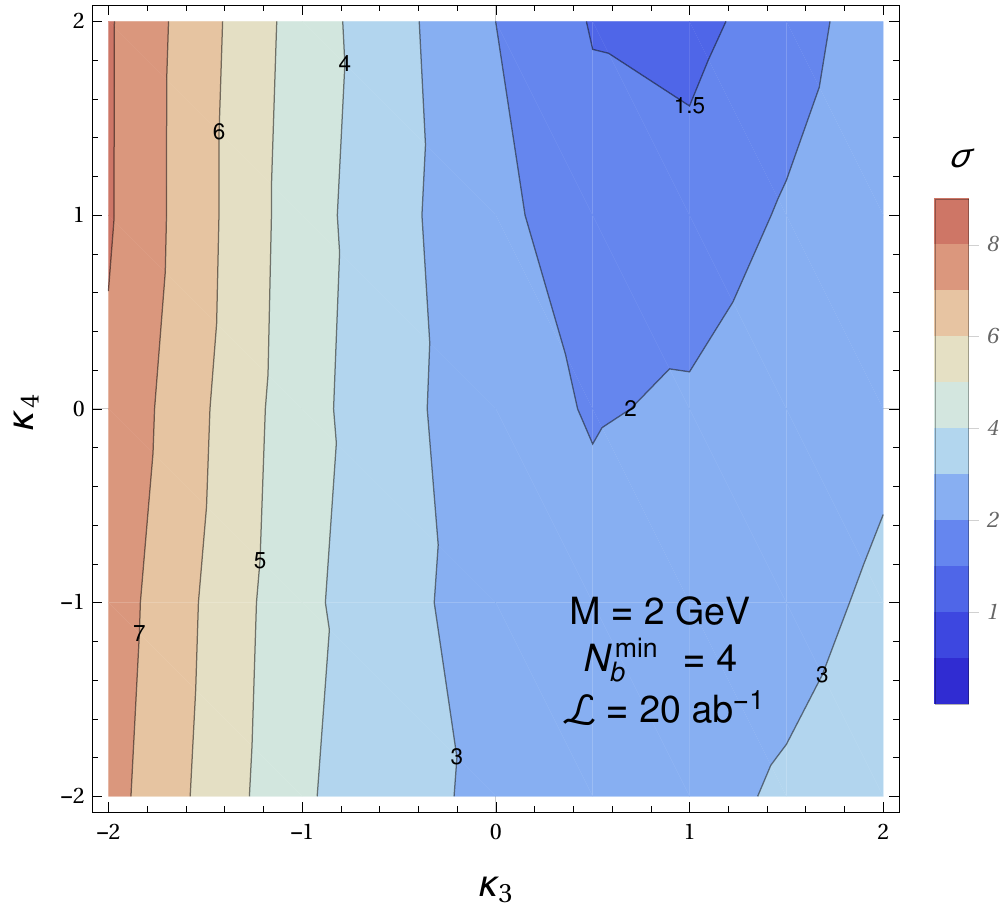}\\[.2cm]
\includegraphics[scale=0.8]{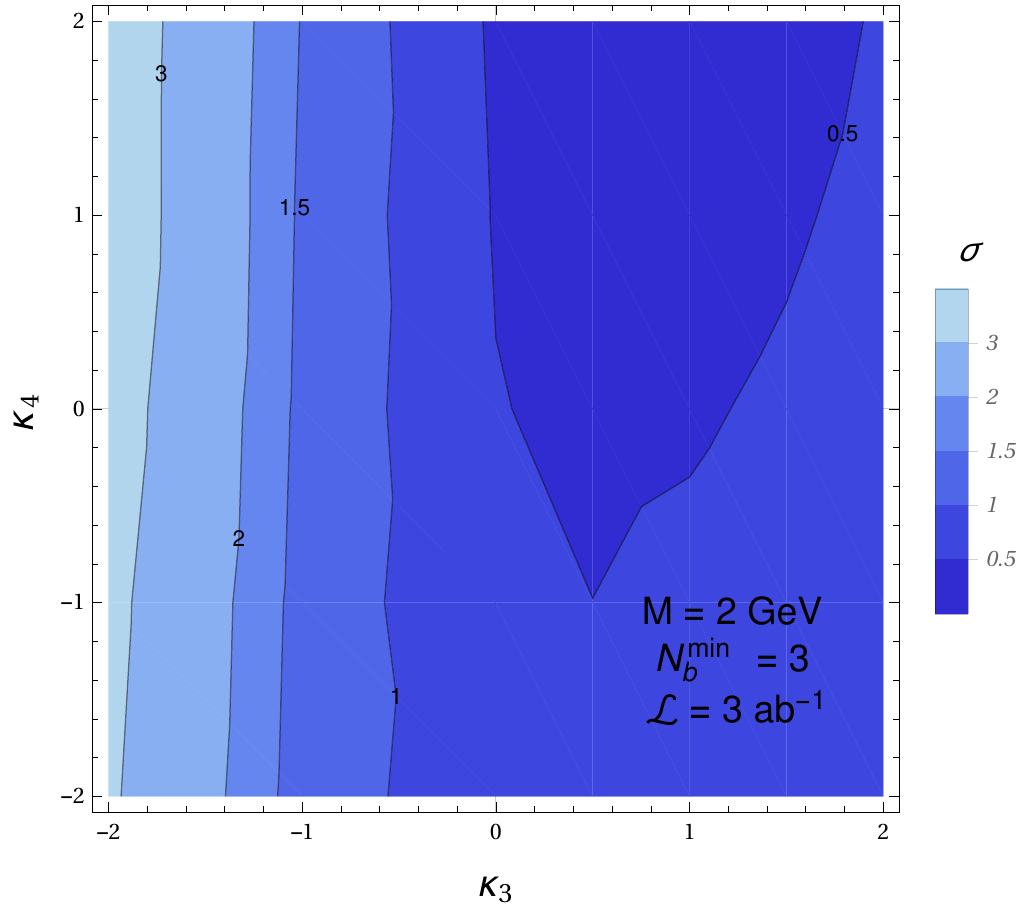}
\includegraphics[scale=0.8]{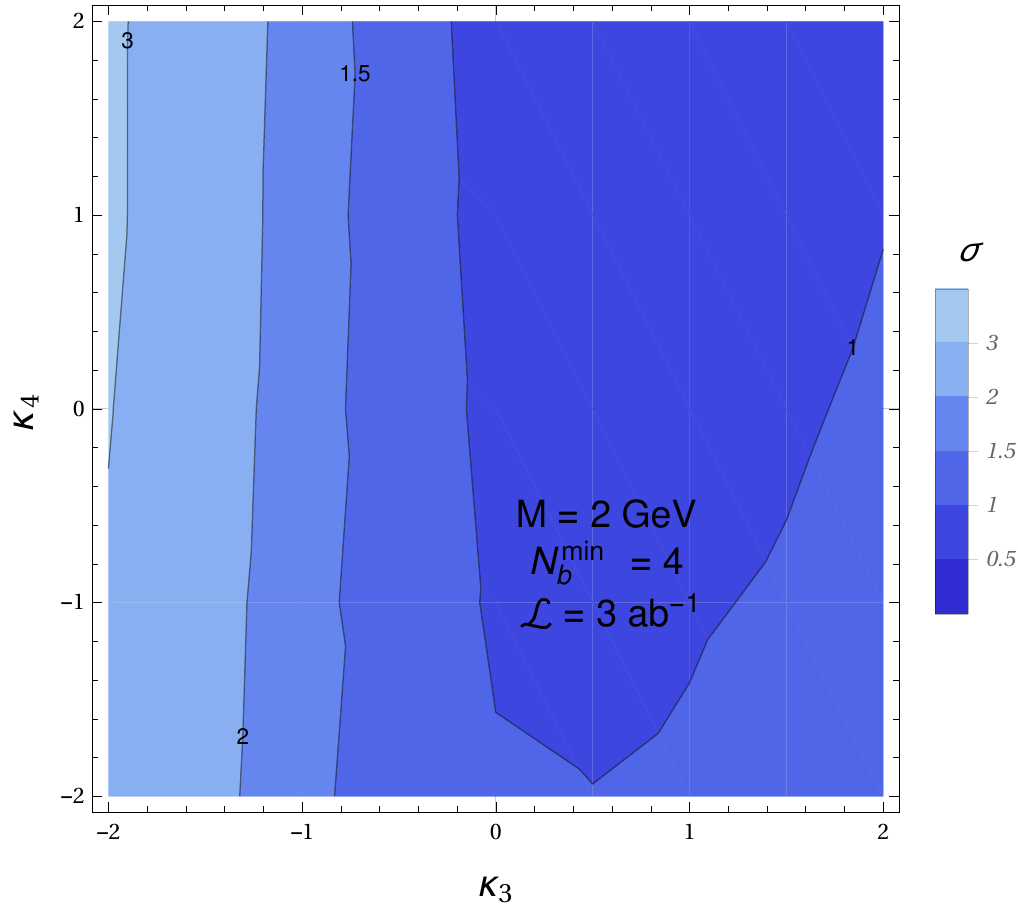}\\[.2cm]
\includegraphics[scale=0.8]{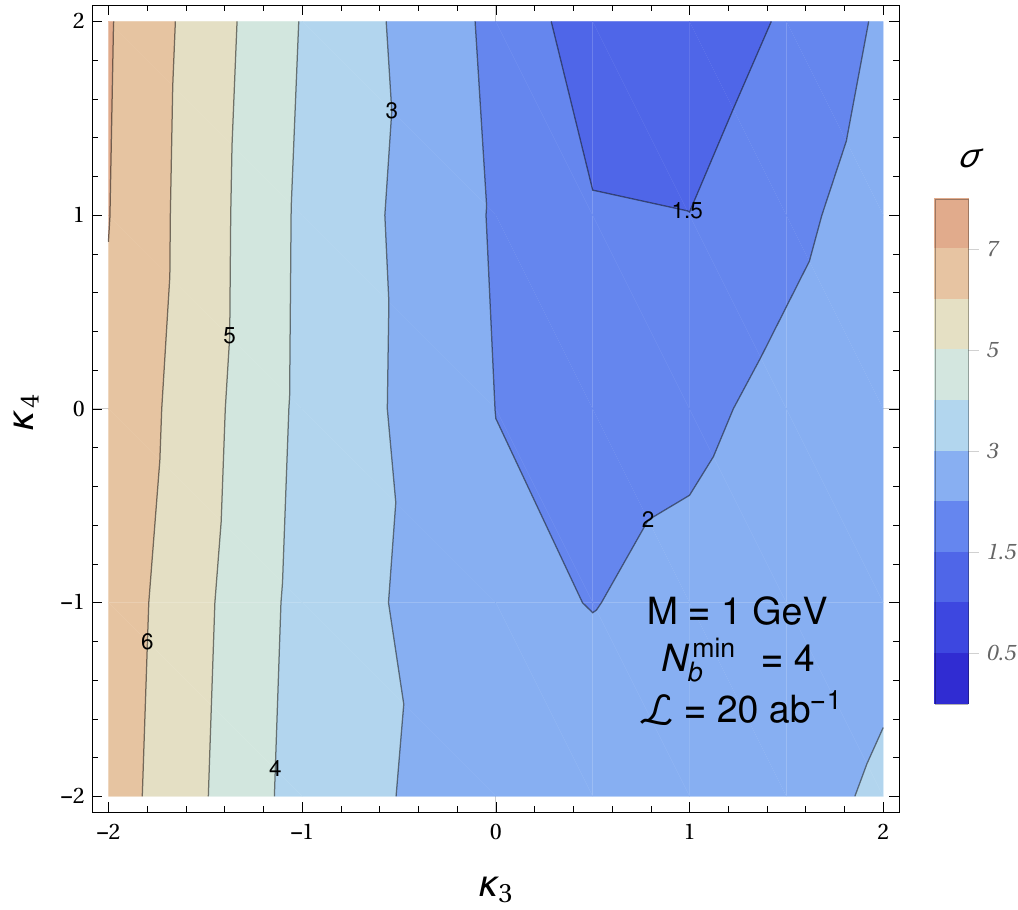}
\includegraphics[scale=0.8]{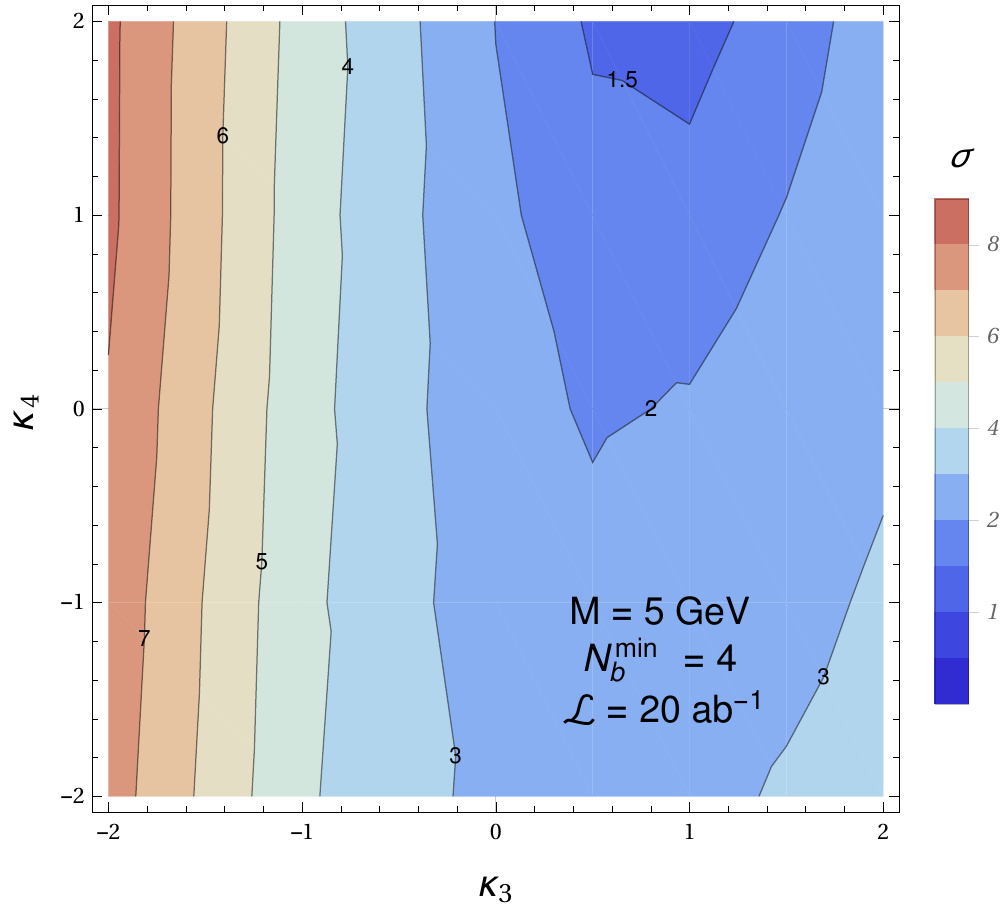}
\caption{Sensitivity of the FCC to the production of a triple-Higgs system
decaying into a $\gamma \gamma b \bar{b} b \bar{b}$ final state when the
selection strategy depicted in the text is followed. We vary the requirement on the
minimum number of $b$-tagged jets $N_b^{\rm min}$, the diphoton mass resolution
$M$ and the luminosity ${\cal L}$ as indicated on the figures.} \label{fig:aExclusionMap2}
\end{figure*}

We have then studied the stability of these conclusions when varying both
$\kappa$ parameters. We have found that in general, the signal efficiency shows
a strong dependence on $\kappa_3$, in particular due to the jet and photon $p_T$
distributions that are slightly harder when $\kappa_3$ is large and positive. On
the contrary, it is less sensitive to $\kappa_4$. Both these conclusions are
illustrated in Figure~\ref{fig:aExclusionMap2} for different configurations of
the $M$ and $N_b^{\rm min}$ variables of the selection strategy, and for two
luminosity goals of 3 and 20~ab$^{-1}$.

Although the constraints on the invariant mass of the three reconstructed Higgs
bosons allow for a good reduction of the background, the signal stays invisible
without invoking $b$-tagging requirements, as illustrated on
Table~\ref{tab:aCutflow} for a given $(\kappa_3, \kappa_4)$ setup. Starting with
a fixed value of \mbox{$M=2$~GeV} \mbox{($| m_h-m_{\gamma\gamma}| < 2$~GeV)} and
aiming towards a luminosity of 20~ab$^{-1}$ of proton-proton collisions at a
center-of-mass energy of 100~TeV, we vary the criterion on the minimum number of
$b$-tagged jets on the first line of
Figure~\ref{fig:aExclusionMap2}. We do not show results for $N_b^{\rm min}=2$ as
this choice does not allow to get a $2\sigma$ significance anywhere on the
probed regions of the parameter space. In contrast, we present the dependence of
$\sigma$ for the cases in which $N_b^{\rm min}=3$ (first line, left panel) and 4 (first
line, right panel). We observe that a good fraction of the parameter space is
covered at the $3\sigma$ level for negative $\kappa_3$ values, and that the
Standard Model case of $(\kappa_3,\kappa_4) = (0,0)$ is even almost reachable at
the $3\sigma$ level for $N_b^{\rm min}=4$. On the second line of the figure, we
show that a lower luminosity phase of the FCC may only be sensitive to a
triple-Higgs signal for extreme deviations from the Standard Model.
On the last line of Figure~\ref{fig:aExclusionMap2}, we fix $N_b^{\rm min}=4$
and vary the value of $M$, \textit{i.e.}, the resolution on the diphoton
invariant mass. We show that resolution choices of 2~GeV (first line, right
figure) or 5~GeV (last line, right figure) lead to a similar sensitivity.
However, adopting a resolution of 1~GeV (last line, left
figure) worsen the situation due to a too low signal efficiency. It however
remains to investigate how a 60\% $b$-tagging performance could be reached at an FCC
(for mistagging rates as low as 1.8\% and 0.1\% for $c$ and lighter jets) and
how it could be feasible to measure a diphoton invariant-mass spectrum at the 2~GeV
level. Additionally, parton-shower, hadronization and underlying event
effects could play a non-negligible role on photon isolation, and the
contributions stemming from the multijet background could impact the significance.
This has been partly studied in Ref.~\cite{Papaefstathiou:2015paa} and seems to be
under good control so that all major effects are covered by the efficiency curves
of Figure~\ref{fig:resolution}. A complete study aiming to design an optimal
analysis strategy, possibly also relying on the simulation of the pile-up, is
left for future work.

In all studied setups in terms of $N_b^{\rm min}$, $M$ and luminosity, we also
observe that with our selection, the significance isolines follow the cross
section (see Figure~\ref{fig:k3_k4_Xsection}), and that it will be
challenging to get any sensitivity to the $\kappa_{3,4} > 0$ territory even
during the high-luminosity phase of the FCC.

On the second panel of Table~\ref{tab:aCutflow}, we investigate the effects of a more
efficient $b$-tagging algorithm (70\%) that also features larger mistagging
rates of 18\% and 1\% for $c$ and lighter jets. Although the signal efficiency is
larger in this configuration, the background contamination is even larger, such that
a poorer significance is yielded.

\subsection{The $h h h \to \tau^+ \tau^- b \bar{b} b \bar{b}$ final state}
\label{Sec2tau4b}
We now consider a triple-Higgs boson signature where the final state is
comprised of four $b$-jets and a pair of hadronically-decaying tau leptons. This
channel has the advantage, compared to the $\gamma \gamma b \bar{b} b \bar{b}$
one, to be associated with a larger branching fraction, but it
receives a more severe background contamination. As in the previous
section, event preselection is performed on the basis of the final state
topology. We demand that all selected events contain at least two tagged
hadronic taus whose transverse momentum $p_T^{\rm \tau_j}$ and pseudorapidity
$\eta^{\rm \tau_j}$ (with \mbox{$j=1,2$}) satisfy
\mbox{$p_T^{\rm \tau_1} > 35$~GeV},
\mbox{$p_T^{\rm \tau_2} > 15$~GeV} and $|\eta_{\rm \tau_j} |< 2.5$,
respectively, and four jets such that \mbox{$p_T^{\rm j_1} > 50$~GeV},
\mbox{$p_T^{\rm j_2} > 30$~GeV}, \mbox{$p_T^{\rm j_3} > 20$~GeV},
\mbox{$p_T^{\rm j_4} > 15$~GeV} and \mbox{$|\eta^{\rm j_i} |< 2.5$} for
\mbox{$i=1,2,3,4$}.

\begin{figure}
\includegraphics[scale=0.8]{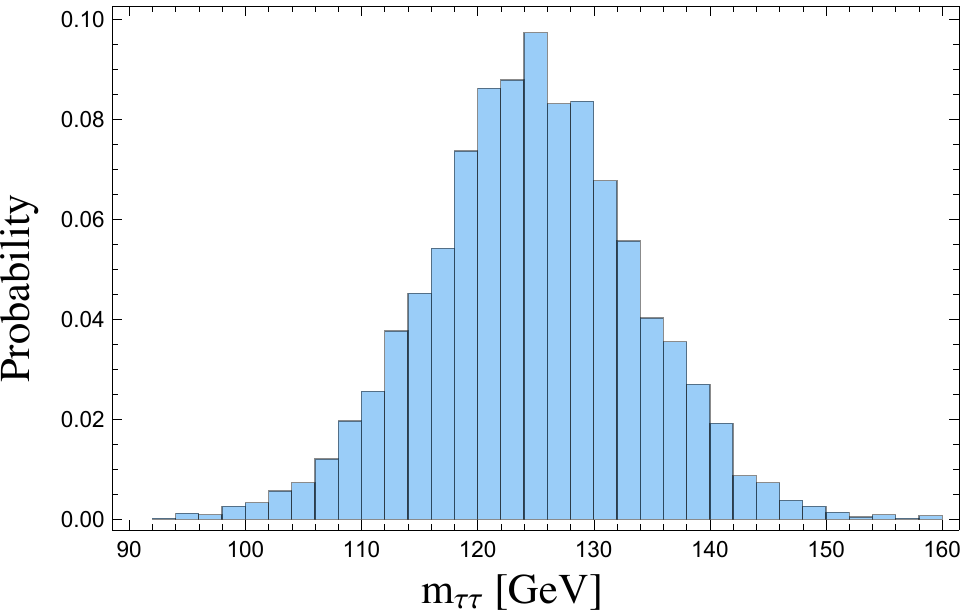}
\caption{Invariant-mass distribution of the Higgs-boson that has been
  reconstructed from the ditau system, after applying the preselection
  strategy. We have considered a benchmark scenario in which $\kappa_3 = -2$ and
  $\kappa_4 = 0$.}
\label{fig:Mh_2ta}
\end{figure}

We then reconstruct two Higgs bosons out of four (possibly fake) $b$-jets using
the method that has been introduced in Section~\ref{Sec2a4b}, and require that
the two reconstructed invariant masses $m_{\rm j j_k}$ are compatible with the
Higgs-boson mass, \mbox{$| m_h - m_{\rm j j_k}| < 15$~GeV} for \mbox{$k=1,2$}.
We additionally impose a constraint on the third reconstructed Higgs boson by
asking the ditau invariant mass $m_{\tau\tau}$ to lie in a 10~GeV window
centered on the Higgs mass. As illustrated in Figure~\ref{fig:Mh_2ta} for a
scenario in which $\kappa_3=-2$ and $\kappa_4 = 0$, this last requirement allows
for the selection of a significant fraction of the signal events.

After additionally demanding that the selected events contain at least
$N_b^{\rm min}$ $b$-jets with $N_b^{\rm min}=3$ or
4\footnote{We have verified that fixing $N_b^{\rm min}$ to 2 was not allowing
us to get any sensitivity to a triple-Higgs signal.}, the Standard
Model background turns out to be dominated by $\tau \tau t\bar{t}$ events where
both top quarks decay hadronically and ditau plus jets events where at least two
jets are $b$-tagged. Moreover, $t\bar{t} h$ and $W^{+} W^{-} b\bar{b}b\bar{b}$
production also contribute significantly, the tau leptons originating here from
top-quark and $W$-boson decays respectively. We finally ignore QCD multijet
background contributions from our analysis, as they are made negligible after
fixing $N_b^{\rm min}$ to either 3 or 4. The corresponding rejection factor,
obtained from parton-level simulations,
is indeed of $10^{10}-10^{12}$. A more precise estimate however
necessitates to include QCD effects such as parton showers and hadronization, and
pile-up. This is left for future
work and we expect, as in the $\gamma\gamma b\bar bb \bar b$ case, that the bulk of
the effects is covered by our mistagging rate parameterization and resolution
functions.

\begin{table*}
\setlength{\tabcolsep}{1.2mm}
\renewcommand{\arraystretch}{1.2}
\scalebox{.82}{\begin{tabular}{c|ccccccccc|c}
 Selection step & Signal  &$\tau \tau b\bar{b}jj$&$\tau \tau b\bar{b}b\bar{b}$
    & $\tau \tau Z_{bb}jj$& $\tau \tau Z_{bb}bb$ &$\tau \tau t\bar{t}$
    & $t\bar{t} h$ & $t\bar{t} z$ &$W^{+} W^{-} b\bar{b}b\bar{b}$ & $\sigma$\\
 \hline \hline
  Preselection & 165~ab & $1.2 \times 10^{7}$~ab & $5.7 \times 10^{4}$~ab
    &$7.4 \times 10^{4}$~ab &$2.8 \times 10^{3}$~ab &$2.1 \times 10^{5}$~ab
    &$7.5 \times 10^{4}$~ab &$1.0 \times 10^{4}$~ab &$4.1 \times 10^{5}$~ab
    &0.21\\
  $| m_h - m_{\rm j j_1, j j_2}| < 15$~GeV &125~ab & $4.2 \times 10^{5}$~ab 
    &$2.3 \times 10^{3}$~ab  &$2.8 \times 10^{3}$~ab &120~ab &
    $2.8 \times 10^{4}$~ab  &$9.5 \times 10^{3}$~ab & 300~ab &
    $1.7 \times 10^{4}$~ab  &0.81\\
  $|m_h - m_{ \tau \tau}| < 10$~GeV & 94~ab & $3.5 \times 10^{3}$~ab
     & 31~ab & 100~ab &7.9~ab & $1.2 \times 10^{4}$~ab  &900~ab 
     & 22~ab & $1.2 \times 10^{3}$~ab & 3.15\\ \hline
  $N_b^{\rm min}=3$ ($\epsilon_b = 0.7$) & 61~ab & 35~ab &  20~ab & 1.0~ab
     &5.1~ab & 520~ab & 590~ab &  14~ab & 770~ab & 6.1\\
  $N_b^{\rm min}=3$ ($\epsilon_b = 0.6$) & 45~ab & 2.6~ab & 15~ab & 0.074~ab
     &3.7~ab & 38~ab & 430~ab & 11~ab & 560~ab & 6.0\\ \hline
  $N_b^{\rm min}=4$ ($\epsilon_b = 0.7$)& 23~ab& 0.17~ab &  7.5~ab
     &$5.0 \times 10^{-3}$~ab & 1.9~ab & 14~ab & 220~ab &  5.3~ab & 280~ab
     & 4.3\\
  $N_b^{\rm min}=4$ ($\epsilon_b = 0.6$) & 12~ab & $1.3 \times 10^{-3}$~ab
     &4.1~ab &$3.7 \times 10^{-5}$~ab  & 1.0~ab & 0.11~ab & 120~ab &  2.9~ab
     &150~ab & 3.2 \\
\end{tabular}}\\[.2cm]
\scalebox{.82}{\begin{tabular}{c|ccccccccc|c}
 Selection step & Signal  &$\tau \tau b\bar{b}jj$&$\tau \tau b\bar{b}b\bar{b}$
    & $\tau \tau Z_{bb}jj$& $\tau \tau Z_{bb}bb$ &$\tau \tau t\bar{t}$
    & $t\bar{t} h$ & $t\bar{t} z$ &$W^{+} W^{-} b\bar{b}b\bar{b}$ & $\sigma$\\
 \hline \hline
  Preselection &68~ab &$5.0 \times 10^{6}$~ab   & $2.4 \times 10^{4}$~ab
    &$3.1 \times 10^{4}$~ab  &$1.2 \times 10^{3}$~ab &$9.2 \times 10^{4}$~ab
    &$3.1 \times 10^{4}$~ab & $4.3 \times 10^{3}$~ab  &$1.7 \times 10^{5}$~ab
    &0.13\\
  $| m_h - m_{\rm j j_1, j j_2}| < 15$~GeV &52~ab &  $1.7 \times 10^{5}$~ab
    &970~ab &$1.2 \times 10^{3}$~ab &48~ab &$1.2 \times 10^{4}$~ab
    &$3.9 \times 10^{3}$~ab & 130~ab & $7.0 \times 10^{3}$~ab &0.52\\
  $| m_h - m_{ \tau \tau}| < 10$~GeV & 39~ab &  $1.5 \times 10^{3}$~ab & 13~ab
    & 43~ab &3.3~ab & $5.1 \times 10^{3}$~ab  &370~ab &  9.1~ab & 490~ab
    & 2.0\\ \hline
  $N_b^{\rm min}=3$ ($\epsilon_b = 0.7$) & 25~ab & 14~ab & 8.6~ab & 0.42~ab
    &2.1~ab & 230~ab & 240~ab &  6.0~ab & 320~ab & 3.9\\
  $N_b^{\rm min}=3$ ($\epsilon_b = 0.6$) & 18~ab& 1.0~ab & 6.3~ab & 0.031~ab
    &1.5~ab & 16~ab & 180~ab &  4.3~ab & 230~ab & 3.9\\ \hline
  $N_b^{\rm min}=4$ ($\epsilon_b = 0.7$) & 9.3~ab &0.071~ab & 3.2~ab
    &$2.1 \times 10^{-3}$~ab & 0.78~ab & 6.3~ab & 90~ab & 2.2~ab & 120~ab
    & 2.8\\
  $N_b^{\rm min}=4$ ($\epsilon_b = 0.6$) & 5.0~ab & $5.2 \times 10^{-4}$~ab
    & 1.7~ab &$1.5 \times 10^{-5}$~ab & 0.42~ab & 0.046~ab & 48~ab &  1.2~ab
    & 63~ab & 2.1\\
\end{tabular}}
\caption{Effects of our selection strategy for an illustrative benchmark
  scenario in which \mbox{$\kappa_3=-2$} and \mbox{$\kappa_4=0$}. We show the
  resulting background and signal cross sections after each of the selection
  steps, together with the related significance that has been calculated for a
  luminosity of 20~ab$^{-1}$. In the upper
  (lower) table, we assume a tau-tagging efficiency of $80\%$ ($50\%$) and a
  mistagging rate of jets as taus of $0.1\%$ ($1\%$).}
\label{tab:TauCutflow}
\end{table*}

\begin{figure*}
\includegraphics[scale=0.8]{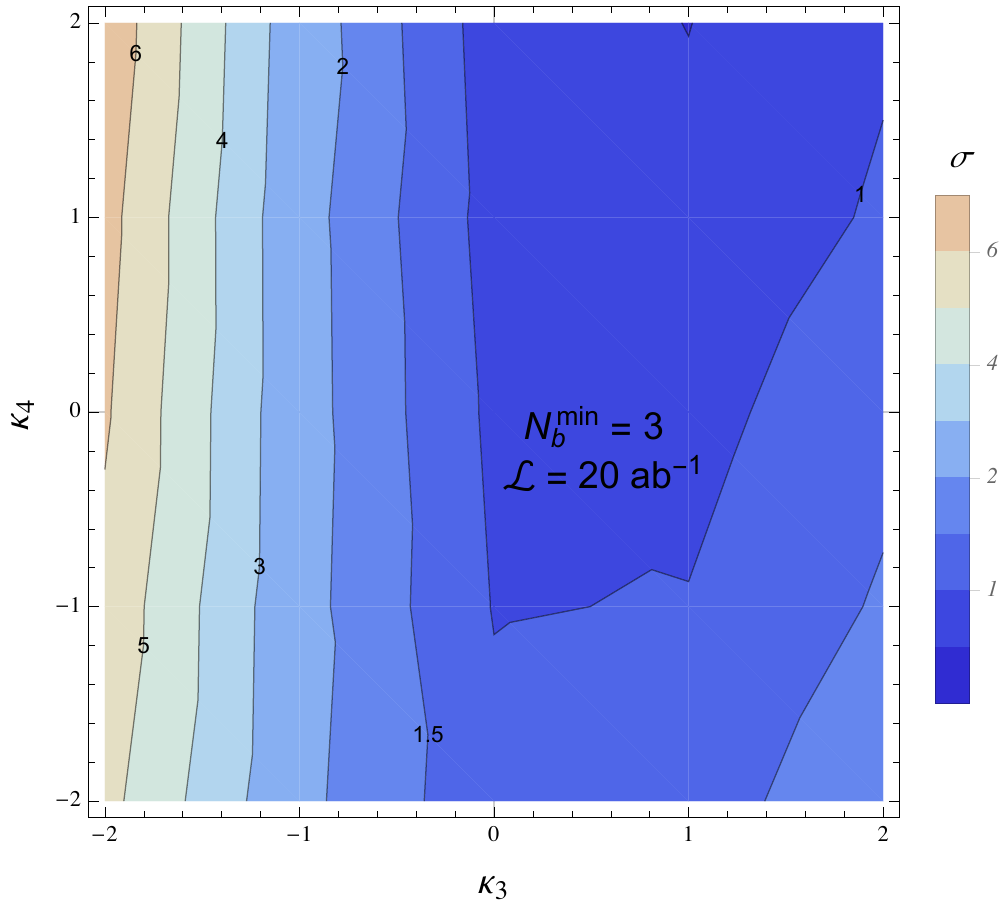}
\includegraphics[scale=0.8]{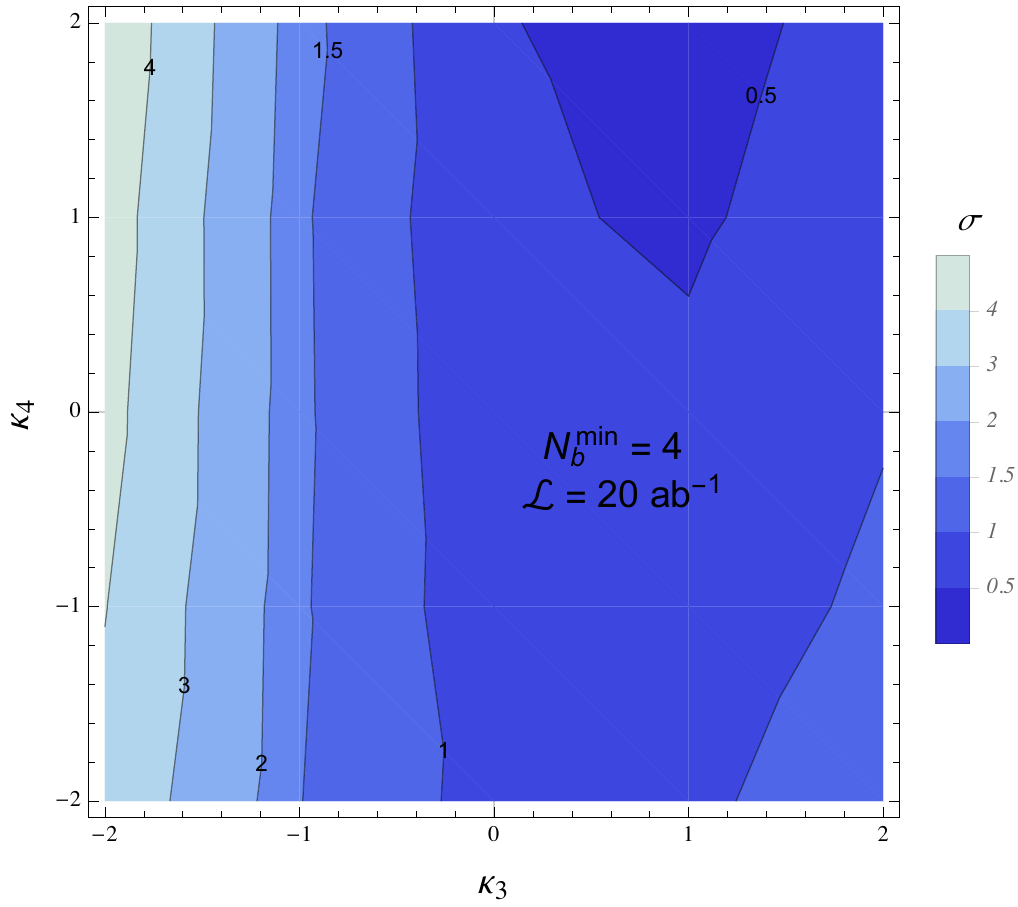}
\caption{Sensitivity of the FCC to the production of a triple-Higgs system
decaying into a $\tau^+\tau^- b \bar{b} b \bar{b}$ final state when the
selection strategy depicted in the text is followed. We show results
for $N_b^{\rm min}=3$ (left) and 4 (right).}
\label{fig:TauExclusionMap}
\end{figure*}

In Table~\ref{tab:TauCutflow}, we present the impact of our selection, for
several $b$-tagging and tau-tagging performances, on both the different
components of the background and the signal. For the latter, we consider a
benchmark scenario in which $\kappa_3 = -2$ and $\kappa_4 = 0$ and the FCC
sensitivity has been calculated assuming an integrated luminosity of
20~ab$^{-1}$ and as in Eq.~\eqref{eq:sign}. We observe that
increasing the minimum number of demanded $b$-tagged jets to $N_b^{\rm min}= 4$
worsens the significance $\sigma$ as the gain in the background rejection is
accompanied with an important suppression of the signal. This is further
depicted in Figure~\ref{fig:TauExclusionMap} where we show the dependence of the
significance on the $\kappa_3$ and $\kappa_4$ parameters when at least three
(left panel) and four (right panel) $b$-tags are requested, for a $b$-tagging
efficiency of $\epsilon_b=$70\% and a mistagging rate of $c$-jets and lighter
jets as $b$-jets of 18\% and 1\%, respectively. Contrary to what has been found
in the diphoton plus four $b$-jets study of Section~\ref{Sec2a4b}, the nature of
the main background contributions is such that using a less efficient
$b$-tagging algorithm with a smaller fake rate is reducing the triple-Higgs
sensitivity.

On the lower panel of the table, we present results in which the
tau-tagger performances are more conservative, with a tagging efficiency of 50\%
for a fake rate of 1\%, and show that for the benchmark scenario under
consideration, one obtains a considerable reduction of the significance. This
feature holds over the entire parameter space so that the possibility of using
the ditau plus four $b$-jets triple-Higgs channel strongly relies on the
availability of an extremely good tau-tagger.

\section{Discussion and FCC design considerations}\label{sec:conclusions}
The form of the Higgs potential $V_{\rm h}$ belongs to the untested parts of the
Standard Model Lagrangian, and it must hence be experimentally probed in the future
to fully unravel the nature of the electroweak symmetry breaking mechanism.
While in the Standard Model, all the parameters driving the bilinear, trilinear
and quartic terms of $V_{\rm h}$ can be fully deduced from the measurement of
the Higgs-boson mass and the value of the vacuum expectation value of the Higgs
field (that is determined from the $W$-boson mass measurement), direct and
independent measurements of all of these parameters can only be achieved with
the study of multiple Higgs-boson production. The prospects for double Higgs
production, that is sensitive to the trilinear Higgs self-coupling, have been
relatively well studied in the context of both the high-luminosity run of the
LHC and the future colliders~\cite{Glover:1987nx,Dawson:1998py,Plehn:1996wb,%
Pierce:2006dh,%
Arhrib:2009hc,Asakawa:2010xj,Grober:2010yv,Contino:2012xk,Dolan:2012rv,%
Gillioz:2012se,Cao:2013si,Nhung:2013lpa,Ellwanger:2013ova,Liu:2013woa,%
No:2013wsa,Baglio:2012np,Shao:2013bz,%
Maltoni:2014eza,Baglio:2014nea,Hespel:2014sla,%
deFlorian:2015moa,Grigo:2015dia,Baur:2002rb,Baur:2003gpa,Baur:2003gp,%
Plehn:2005nk,Binoth:2006ym,Chen:2014xra,Bhattacherjee:2014bca,Dawson:2015oha}.
Although the knowledge of the trilinear Higgs
self-coupling is not enough to fully test the Standard Model nature of the Higgs
potential, triple Higgs-boson production, that is sensitive to both the
trilinear and quartic Higgs self-couplings, remains less explored so
far~\cite{Papaefstathiou:2015paa,Chen:2015gva}.

In this paper, we have continued to fill this gap and studied the prospects for
measuring all renormalizable interaction strengths of the Higgs potential at a
future proton-proton collider running at a center-of-mass energy of 100~TeV. We
have focused on the production of a triple-Higgs-boson system and examined two
of its specific signatures. More precisely, we have considered final states
comprised of four $b$-jets, and either a pair of photons or a pair of tau
leptons. We have furthermore decided to be agnostic of any specific assumption
on the future collider detector capacities and provided instead guidelines for
detector designs that would allow for the observation a triple-Higgs signal. For the
same reason, the investigation of channels that are associated with large
branching fractions, such as the six $b$-jet or the four $b$-jet plus a
$W$-boson pair modes, but whose analysis requires a deeper knowledge of the FCC
detector performances, has been left for a future work.

Our study indicates that triple-Higgs production is more sensitive to new
physics contributions to the trilinear Higgs self-coupling (collected under the
$\kappa_3$ parameter in our theoretical model description) than to the quartic
one (collected under the $\kappa_4$
parameter). These findings closely follows the dependence of the triple-Higgs
total cross section on the $\kappa_i$ parameters so that the sensitivity reach
of the FCC in the $(\kappa_3, \kappa_4)$ plane mostly extends to regions in
which the trilinear coupling is large and negative. This conclusion holds for
any value of $\kappa_4$.
In order to assess how the FCC would be sensitive to deviations from the
Standard Model in the Higgs self-interactions and to make our results useful for
detector design studies, we have made use of Monte Carlo simulations of both the
signal and the Standard Model background. We have additionally explored the
impact of different $b$-tagging and tau-tagging performances, as well as of
different diphoton invariant-mass resolutions.

We have chosen two $b$-tagging setups with efficiencies of 70\% and 60\%,
respectively, for related mistagging rates of a $c$-jet (light jet) as a $b$-jet
of 18\% (1\%) and 1.8\% (0.1\%). We have found that the best expectation is
obtained by requiring at least three $b$-tagged jets (\mbox{$N_b^{\rm min}=3$})
in the $\tau^+ \tau^- b \bar{b} b \bar{b}$ and four $b$-tagged jets
(\mbox{$N_b^{\rm min}=4$}) in the $\gamma \gamma b \bar{b} b \bar{b}$ mode.
These choices indeed allow both for an efficient background rejection and to
maintain a high signal efficiency (at the 50\% level). Due to the different
natures of the dominant components of the background and the $b$-tagging
requirements of both analysis strategies, the four $b$-jets plus diphoton study
has been found to exhibit better results with a small fake rate (at the
price of a smaller $b$-tagging efficiency), which contrasts with the four
$b$-jets plus a ditau channel for which it is better to make use of a more
efficient $b$-tagging algorithm (exhibiting thus a larger fake rate).
We have adopted two benchmark tau-tagging performance setups. The first one
is optimistic and features a tagging efficiency of 80\% for an associated
mistagging rate of a jet as a tau of 0.1\%. The second setup is more
conservative, the tagging efficiency being of 50\% and the fake rate
of 1\%. We have found that only a very efficient tau-tagging algorithm
provides hopes for the four $b$-jets plus a tau pair channel to be
sensitive to a triple-Higgs boson signal. In this case, this channel can be
almost as competitive, and thus complementary, to the four $b$-jets plus two
photons one. Nevertheless, such optimistic tau-tagging performances certainly need to be
assessed by an experimental study, while our work warrants some benefits for
an improved tau-tagging at future colliders.
We have finally investigated the effects of different diphoton mass resolutions
for the four $b$-jets plus a photon pair channel, imposing the reconstructed
Higgs-boson mass from the diphoton system to be compatible with the true
Higgs-mass at the 1, 2 and 5~GeV level. For the first choice
(\mbox{$|m_h-m_{\gamma\gamma}|<1$~GeV}), we have observed that the related
reduced signal efficiency was worsening the FCC sensitivity to the triple-Higgs
signal, while the two other cases are implying improved results, with a smaller
mass resolution being preferred.

\begin{figure}
\includegraphics[scale=0.8]{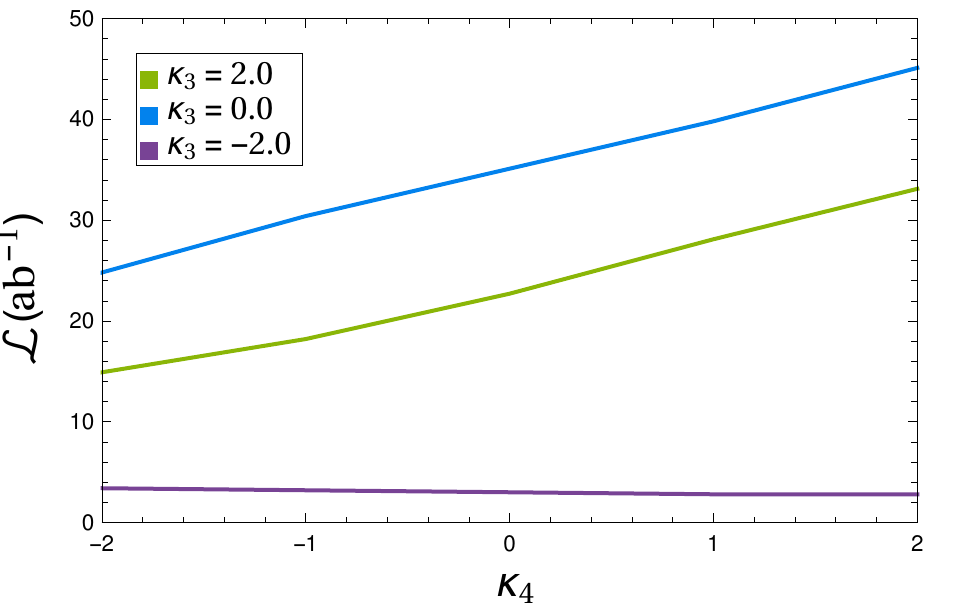}
\caption{Minimum FCC luminosities that are required to achieve a $3\sigma$
  sensitivity to a triple-Higgs signal in terms of the $\kappa_4$ parameter for
  fixed values of $\kappa_3$.}
\label{fig:LuminosityGoal}
\end{figure}

All our results mainly rely on a
20~ab$^{-1}$ FCC luminosity (unless stated otherwise). We have however studied
departures from this choice in order to get a useful ground for estimating
luminosity goals of a future 100~TeV hadron colliders that would allow for the
measurement of the quartic Higgs self-coupling. Focusing on the most promising
analysis strategy in which the $\gamma \gamma b \bar{b} b \bar{b}$ channel is
used with a diphoton invariant mass requirement of
\mbox{$| m_h - m_{ \gamma \gamma}| < 2$~GeV} and a demand of at least four
$b$-tagged jets, we show in Figure~\ref{fig:LuminosityGoal} the minimum FCC
luminosities that would be required to achieve a $3\sigma$ sensitivity in terms
of the value of the $\kappa_4$ parameter and for several fixed values of
$\kappa_3$. As very large and negative $\kappa_3$ values ensure an important
enhancement of the triple-Higgs production cross section, a $3\sigma$
observation of a triple-Higgs signal is guaranteed with a few ab$^{-1}$
regardless of the size of the new physics contributions to the Higgs quartic
coupling. For larger values of $\kappa_3$, a few tens of ab$^{-1}$ (which
roughly corresponds to a period of 20-30 years of FCC
running~\cite{FCC,FCCihep}) are required, so that
one will get sensitivity to only a fraction of the scanned
region of the $(\kappa_3,\kappa_4)$ parameter space.

\bigskip
\emph{Acknowledgements:} 
We thank the HTCaaS group of the Korea Institute of Science and
Technology Information (KISTI) for providing the computing resources that have
allowed for the completion of this project. BF and SL are grateful to the Aspen
Center for Physics for its hospitality while part of this work was initiated,
and we also acknowledge the Korea Future Collider Study Group (KFCSG) for
motivating us to proceed with this work. JHK is supported by the IBS Center for
Theoretical Physics of the Universe and Center for Axion and Precision Physics
Research (IBS-R017-D1-2015-a00). SL has been supported in part by the National
Research Foundation of Korea(NRF) grant funded by the Korea government(MEST)
(NRF-2015R1A2A1A15052408).

\bibliography{draft_paper}
\end{document}